# Nuclear structure and the nucleon effective mass: explorations with the versatile KIDS functional[‡]

P. Papakonstantinou[1,*], H. Gil[2]

[1] *Rare Isotope Science Project, Institute for Basic Science, Daejeon 34047, Korea*
[2] *Department of Physics, Kyungpook National University, Daegu 41566, Korea*

**Abstract**    The connection from the structure and dynamics of atomic nuclei (finite nuclear system) to the nuclear equation of state (thermodynamic limit) is primarily made through nuclear energy-density functional (EDF) theory. Failure to describe both entities simultaneously within existing EDF frameworks means that we have either seriously misjudged the scope of EDF or not fully taken advantage of it. Enter the versatile KIDS Ansatz, which is based on controlled, order-by-order extensions of the nuclear EDF with respect to the Fermi momentum and allows a direct mapping from a given, immutable equation of state to a convenient Skyrme pseudopotential for applications in finite nuclei. A recent proof-of-principle study of nuclear ground-states revealed the subversive role of the effective mass. Here we summarize the formalism and previous results and present further explorations related to giant resonances. As examples we consider the electric dipole polarizability of $^{68}$Ni and the giant monopole resonance (GMR) of heavy nuclei, particularly the fluffiness of $^{120}$Sn. We find that the choice of the effective mass parameters and that of the compression modulus affect the centroid energy of the GMR to comparable degrees.

## INTRODUCTION

Nuclear structure data provide access to basic information about the nuclear equation of state. Nuclear masses, charge radii, and giant resonances have been analyzed over the years to estimate the saturation point of cold symmetric matter, its nuclear compression modulus, and its symmetry energy and slope. The connection between the structure and dynamics of atomic nuclei (finite nuclear system) and the nuclear equation of state (thermodynamic limit) is primarily made through nuclear energy-density functional (EDF) theory [1,2]. Failure to describe both within existing EDF frameworks would imply either that we have seriously misjudged the scope of EDF or a more optimistic scenario: perhaps the full potential of EDFs has not been exploited.

The versatile KIDS Ansatz involves a controlled, order-by-order expansion and extension of the nuclear EDF with respect to the Fermi momentum [3] and allows a direct mapping from a given, immutable equation of state to a convenient Skyrme-type functional of arbitrarily extended form. Using a minimal Skyrme-type form, recent results for nuclear ground-state

---

properties [4] have shown the power of the approach and revealed the subversive role of the effective mass parameters. Here we summarize the formalism and previous findings and we present further explorations related to giant resonances. As examples of current active interest we consider the electric dipole polarizability of $^{68}$Ni and the giant monopole resonance of heavy nuclei, particularly the fluffiness of $^{120}$Sn.

**KIDS AND THE NUCLEAR EQUATION OF STATE**

The KIDS energy per particle is written as an expansion in powers of the Fermi energy, or equivalently the cubic root of the density. Any number and order of terms can be considered in principle, but the optimal form has been determined to include up to four terms beyond the free kinetic energy term [3,4] and can be expressed compactly as follows:

$$\mathcal{E}(\rho,\delta) = \sum_{i=-1}^{3} c_i(\delta)\rho^{1+i/3}.$$

The parameters $c_i(\delta)$ are model parameters, except for $c_{-1}(\delta)$ which corresponds to the kinetic energy of a free Fermi gas and is given by [5]

$$c_{-1}(\delta) = \frac{3}{10}\frac{\hbar^2}{2m}\left[(1-\delta)^{\frac{5}{3}} + (1+\delta)^{\frac{5}{3}}\right]\left(\frac{3\pi^2}{2}\right)^{\frac{2}{3}}.$$

In the present work the symmetry energy is defined as the difference between the energy of pure neutron matter (PNM) and the energy of isospin-symmetric nuclear matter (SNM). (Alternatively, one can define it as the derivative with respect to asymmetry at the saturation point [5,6].) Then it can be expressed as

$$S(\rho) = \sum_{i=-1}^{3}[c_i(1) - c_i(0)]\rho^{1+i/3} \equiv \sum_{i=-1}^{3}\beta_i\rho^{1+i/3}.$$

All parameters characterizing the nuclear equation of state at the saturation point assume analytical expressions. Let us define the generic derivatives of the EoS of SNM and of the symmetry energy, respectively,

$$R_{\text{SNM}}^{(n)}(\rho) = (3\rho)^n\frac{d^n\mathcal{E}(\rho,0)}{d\rho^n} \; ; \; R_{\text{sym}}^{(n)}(\rho) = (3\rho)^n\frac{d^nS(\rho)}{d\rho^n}.$$

The incompressibility of SNM is then simply $K_0 = R_{\text{SNM}}^{(2)}(\rho_0)$, etc. Similarly, the parameters characterizing the symmetry energy are defined at the saturation point as its value $J = S(\rho_0) = R_{\text{sym}}^{(0)}(\rho_0)$, its slope $L = R_{\text{sym}}^{(1)}(\rho_0)$, its curvature $K_{\text{sym}} = R_{\text{sym}}^{(2)}(\rho_0)$, its skewness $Q_{\text{sym}} = R_{\text{sym}}^{(3)}(\rho_0)$, and its kurtosis $R_{\text{sym}}^{(4)}(\rho_0)$.

For given (assumed) values of EoS parameters (these include the saturation density of SNM $\rho_0$, the saturation energy $\varepsilon_0$, and its slope which by definition vanishes at saturation, as well as the symmetry energy parameters), an equal number $N$ of KIDS parameters can be determined by solving simple algebraic $N \times N$ systems. This becomes clear if we write out the above equations as follows:

$$\left[R_{\text{SNM}}^{(n)}(\rho_0) - 3^n c_{-1}(0)\frac{\Gamma(\frac{5}{3})}{\Gamma(\frac{5}{3}-n)}\rho_0^{2/3}\right] = \sum_{i=0}^{3} c_i(0)\left[\frac{3^n\Gamma(2+\frac{i}{3})}{\Gamma(2+\frac{i}{3}-n)}\rho_0^{1+i/3}\right],$$

$$\left[R_{\text{sym}}^{(n)}(\rho_0) - 3^n \beta_{-1} \frac{\Gamma(\frac{5}{3})}{\Gamma(\frac{5}{3}-n)} \rho_0^{2/3}\right] = \sum_{i=0}^{3} \beta_i \left[\frac{3^n \Gamma(2+\frac{i}{3})}{\Gamma(2+\frac{i}{3}-n)} \rho_0^{1+i/3}\right],$$

where the known (assumed) quantities have been collected in square brackets while the unknowns are $c_i(0)$ and $\beta_i$ (or $c_i(1) = \beta_i + c_i(0)$) for i ≥ 0. It is implied that the coefficients $\frac{3^n \Gamma(2+\frac{i}{3})}{\Gamma(2+\frac{i}{3}-n)} = (3+i)i(-3+i)\ldots(-3n+6+i)$ vanish if $i = 0,3,\ldots$ (multiple of 3) and $n > 1 + i/3 = 1,2,\ldots$, respectively.

In the first applications of the KIDS functional [3,4,6-9] the following SNM equation of state was adopted:

$$\rho_0 = 0.16 \text{ fm}^{-3}, \varepsilon_0 = -16 \text{ MeV}, K_0 = 240 \text{ MeV}, Q_0 = -373 \text{MeV}.$$

The values of $\rho_0, \varepsilon_0, K_0$ were imposed as empirical, while the value for $Q_0$ was obtained by setting $c_3(0) = 0$. The justification of that choice is discussed in [3]. For the symmetry energy the following values were adopted (rounded values to the second significant digit are shown here):

$$J = 33\text{MeV}, L = 50\text{MeV}, K = -160 \text{ MeV}, Q = 590 \text{ MeV}.$$

They correspond to a fit of $c_i(1)$ to the Akmal-Pandharipande-Ravenhall (APR) EoS for pure neutron matter (PNM) [10]. The resulting parameter set, labeled ad-2 [3], was used with success in the description of finite nuclei, producing results on par with modern, refined Skyrme functionals [4]. It has served also as baseline for further explorations of parameters. In particular, in [6,7], for various values of the curvature and skewness of the symmetry energy, and in addition the skewness of SNM [7], the neutron-matter EoS and the neutron-star mass-radius relation are explored. That ongoing work indicates that high-order parameters of the symmetry energy could be constrained from astronomical observations and from pseudodata (results of ab initio computations) for dilute neutron matter. The corollary is that the lower-order parameters J and L are not the only determinants of the symmetry energy in dense and dilute matter.

**KIDS AND SKYRME FUNCTIONALS: NUCLEAR STRUCTURE**

For applications in nuclei we turn to the Kohn-Sham framework and specifically the formalism of Skyrme-type functionals [1,5]. A Skyrme functional can readily be employed in Hartree-Fock computational codes for calculating nuclear ground-state properties and in random-phase-approximation codes for calculating the properties of collective excitations within linear response theory. In this sense the KIDS functional goes well beyond other recently proposed analytical EoSs like the so-called meta-modelling [11,12].

The minimal required Skyrme pseudopotential has three density-dependent terms, as follows:

$$v_{ij} = (t_0 + y_0 P_s)\delta(\mathbf{r_{ij}}) + \frac{1}{2}(t_1 + y_1 P_s)[\delta(\mathbf{r_{ij}})\mathbf{k}^2 + \text{h.c.}] + (t_2 + y_2 P_s)\mathbf{k}' \cdot \delta(\mathbf{r_{ij}})\mathbf{k}$$
$$+ i W_0 \mathbf{k}' \times \delta(\mathbf{r_{ij}})\mathbf{k} \cdot (\boldsymbol{\sigma}_i - \boldsymbol{\sigma}_j) + \frac{1}{6}\sum_{n=1}^{3}(t_{3n} + y_{3n} P_s)\rho^{\frac{n}{3}}\delta(\mathbf{r_{ij}}).$$

The $t_i, y_i$ parameters have a direct analytical correspondence with the KIDS EoS parameters. In particular, given a KIDS EoS (parameters $c_i(0), c_i(1)$), the Skyrme parameters are given by:

$$t_0 = \frac{8}{3}c_0(0), y_0 = \frac{8}{3}c_0(0) - 4c_0(1); \quad t_{3n} = 16c_n(0), y_{3n} = 16c_n(0) - 24c_n(1) \ (n \neq 2)$$

$$t_{32} = 16c_2(0) - \frac{3}{5}\left(\frac{3}{2}\pi^2\right)^{2/3}\theta_s, \quad y_{32} = 16c_2(0) - 24c_2(1) + \frac{3}{5}(3\pi^2)^{\frac{2}{3}}\left(3\theta_\mu - \frac{\theta_s}{2^{2/3}}\right)$$

with

$$\theta_s \equiv 3t_1 + 5t_2 + 4y_2, \ \theta_\mu \equiv t_1 + 3t_2 - y_1 + 3y_2.$$

The above equations reveal two possible sources of the term proportional to $\rho^{5/3}$ (parameters $t_1, y_1, t_2, y_2 \leftrightarrow c_{32}(0), c_{32}(1)$) and the necessity to either reduce the number of unknowns, e.g. by setting $y_1 = y_2 = 0$, or specify two more independent parameters, most conveniently those related to the isoscalar and isovector effective masses

$$m^*/m = 1 + \frac{m}{8\hbar^2}\rho\theta_s, \ m_v^*/m = 1 + \frac{m}{4\hbar^2}\rho(\theta_s - \theta_\mu).$$

Additionally, a spin-orbit term is necessary. The strength $W_0$ of such a term is unconstrainable from infinite matter and from bulk properties of spin-saturated nuclei. Therefore it is constrained at present through bulk properties (energies and charge radii) of $^{48}$Ca and $^{208}$Pb.

The freedom to vary these parameters independently, absent in conventional fitting protocols, is a great asset for the KIDS functional: It allows us to unambiguously conclude whether a specific observable is sensitive to the choice of the effective mass values when all else is kept constant, namely the bulk EoS parameters.

Let us now recapitulate the procedure for obtaining the Skyrme parameters from the KIDS parameters. The parameters $t_0, y_0, t_{31}, y_{31}, t_{33}, y_{33}$ are readily derived from the above analytical expressions. As already mentioned, the term in the expansion proportional to $\rho^{5/3}$ (parameters $c_2(0), c_2(1)$) can originate in a purely density-dependent pseudopotential of coupling strength proportional to $\rho^{2/3}$ (parameters $t_{32}, y_{32}$) or from the usual momentum dependence of the Skyrme functional, determined by the standard parameters $t_1, t_2, y_1, y_2$. The total term is fixed from the SNM and PNM EOSs; only the partition is unknown. Two ways have been used so far to determine the partition, i.e., constrain the momentum dependence, which we describe below.

A crude way is to set $y_1 = y_2 = 0$ and determine the partition nad hence $t_1, t_2$ from a fit to nuclear data. (A fit to the masses and radii of spin-saturated $^{16}$O and $^{40}$Ca can be done independently of $W_0$ [8].) This procedure generally favors a value for the isoscalar effective mass which is rather close to the bare nucleon mass. One such Skyrme parameter set based on the ad-2 EoS is indicated as "KIDS0" [8]. Very good results were obtained for static properties, but the approximation may be insufficient for dynamics.

A preferable way then is to retain the freedom in the isoscalar and isovector effective masses and use the additional variables to determine all four Skyrme parameters which control the momentum dependence, namely $t_1, t_2, y_1, y_2$. We summarize this second procedure as follows:

- We begin with a given set of eight parameters $c_i(0), c_i(1)$, e.g., the ad-2 set - see previous section. We wish to transpose it to a generalized Skyrme pseudopotential and functional with three density dependent terms. We have 12 bulk parameters to determine, plus the spin-orbit term.
- The 6 parameters $t_0, y_0, t_{31}, y_{31}, t_{33}, y_{33}$ are readily obtained. There are four independent parameters left (combinations of $t_1, y_1, t_2, y_2$).

- We assume a value for the isoscalar effective mass $\mu_s \equiv m^*/m$ and a value for the isovector effective mass $\mu_v \equiv m_v^*/m$, within a physically acceptable range. This fixes $\theta_s, \theta_\mu$, which, together with $c_2(0), c_2(1)$, can fix $t_{32}, y_{32}$. We are left with two independent parameters within (combinations of) $t_1, y_1, t_2, y_2$.
- The two parameters, along with the spin-orbit constant constant $W_0$, are fitted to the masses and charge radii of $^{40}$Ca, $^{48}$Ca, $^{208}$Pb (six data).

Results for the mass, charge radius, and neutron skin thickness of magic nuclei throughout the nuclear chart, based on the ad-2 parameterization, have been presented in [4]. Here we summarize the main findings.

The most significant result is that the nuclear masses and charge radii, which represent bulk and static properties, turn out to be practically independent of the assumed value of the effective mass. In particular, results were obtained with parameters corresponding to $m^*/m = 0.7, 0.8, 0.9, 1.0$ and with KIDS0. The fit quality to the six data and the predictions were practically the same. Moreover, predictions of known data which were not used in the fit were on par with predictions from existing Skyrme functionals obtained by fitting to many data. The only differences observed among the KIDS sets were in predictions for the neutron skin thickness of only specific nuclei, not exceeding current experimental uncertainties.

The conclusion is that bulk and static properties do not provide information on the in-medium effective mass, but they provide a direct link to the bulk EoS. It is then possible to provide predictions for exotic nuclei, based solely on an EoS. In Ref. [4], for example, based on accepted values for the SNM EoS and the APR PNM EoS (ad-2 parameter set), the energy per particle and charge radius of the recently discovered $^{60}$Ca isotope [13] are predicted to be 7.65-7.67 MeV and about 3.64 fm, respectively.

Next we ask whether the same conclusion can be reached for other static properties, such as the electric dipole polarizability, and what the effect is on dynamic properties, such as giant resonances. A selection of preliminary results is presented next.

**RESULTS – I: DIPOLE POLARIZABILITY AND EFFECTIVE MASS**

The electric dipole polarizability $a_D$ of a given nucleus quantifies the nucleus' intrinsic static response to an electric field, or equivalently an isovector field. Therefore it must be largely determined by the symmetry energy, especially on the nuclear surface, where isovector polarization primarily develops. A statistical analysis has revealed that the dipole polarizability is correlated with the neutron-skin thickness of neutron rich nuclei, at least within a given theoretical model [14].

In practice, $a_D$ is calculated from the inverse energy-weighted sum (IEWS) of the dynamic electric-dipole strength function. The same holds for its measurement. A complete photoresponse spectrum is required, at least up to high enough energies that the IEWS is converged. Accurate experiments are nowadays possible for stable nuclei [15]. Here we focus on the neutron-rich $^{68}$Ni, for which measurements exist [16,17], be it with large uncertainties.

The photoresponse, represented by the electric dipole strength function $S_{E1}$, is shown in the left panel of Fig. 1. The calculations correspond to the ad-2 parameter set and different assumptions for the isoscalar effective mass, as specified in the figure. Clearly the fragmentation pattern is influenced by the choice, indicating the role of the underlying shell structure and Landau damping. The position and strength of the structure in the vicinity of 10 MeV, attributable to a collective pygmy resonance, are also affected. The total IEWS though, i.e., the polarizability $a_D$ shows discrepancies of at most 2%. (The neutron skin thickness is found to be 0.19-0.20fm.) The cumulative sum is shown on the right of Fig. 1 and the converged result compares well with the experimental datum.

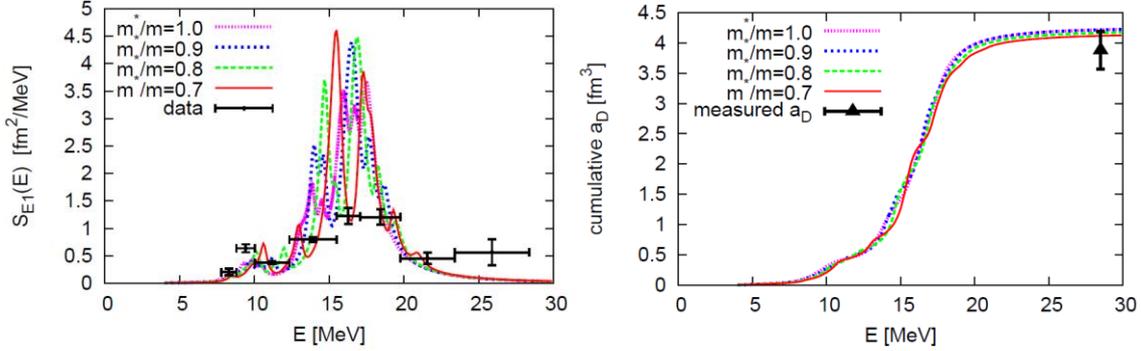

**Fig. 1**. Left: RPA predictions for the electric dipole response of $^{68}$Ni based on the KIDS-ad-2 equation of state with no adjustments (bulk unchanged) and under different assumptions for the isoscalar effective mass, compared with data [15,16]. Right: Respective results for the dipole polarizability (revised datum from [16]).

In these calculations, the isovector effective mass is kept constant. As a result, the isovector enhancement factor and therefore the energy-weighted sum also remain unchanged.

Although it is premature to draw definitive conclusions, a few remarks are in order. The dipole polarizability is almost unaffected by the choice of isoscalar effective mass and the same has been generally found for the neutron skin thickness [4]. On the other hand, details of the $S_{E1}$ distribution, and notably the strength and position of the pygmy dipole resonance, do show some sensitivity. As regards the pygmy resomance region, it is interesting to note in Fig.1 that the IEWS for all four computations equalize between 11 and 12 MeV. Further explorations are underway.

**RESULTS – II: COMPRESSION MODE AND THE EFFECTIVE MASS**

Dynamic properties such as the energies of giant resonances could be influenced by the momentum dependence of the in-medium nuclear interaction, encoded here in the effective mass. As a case study we choose the isoscalar giant monopole resonance (GMR). As a compression mode it is of great interest, because it is expected to provide information on the nuclear incompressibility. Puzzles remain though, for example, the "fluffiness" of the Sn isotopes – by which one refers to the observation that nuclear energy-density functionals which describe accurately the properties of the GMR in various heavy nuclei overestimate the GMR energy in Sn isotopes by about 1MeV [18]. Here we examine to what extent some of the discrepancy can be attributed to the choice of effective mass.

Fig.2 shows the isoscalar monopole strength distribution of $^{208}$Pb and $^{120}$Sn calculated using the KIDS-ad-2 parameter set ($K_0$=240MeV) and a set with the same EoS parameters except for the compression modulus ($K_0$=220MeV). Again various values are adopted for the isoscalar effective mass. The calculated and measured centroids are indicated by triangles and arrows, respectively.

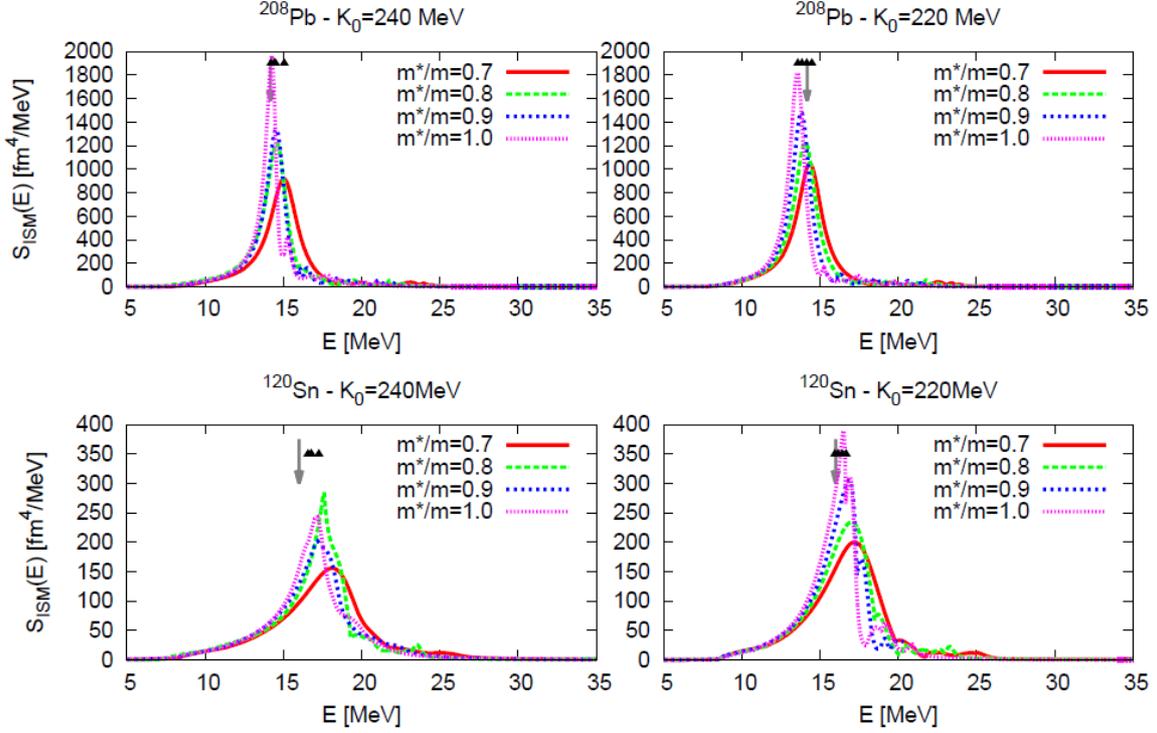

**Fig. 2**. The isoscalar monopole response of $^{208}$Pb (top) and $^{120}$Sn (bottom) based on the KIDS-ad-2 EoS ($K_0$=240 MeV, left) and an EoS with the same parameters except for the incompressibility ($K_0$=220 MeV, right), under different assumptions for the isoscalar effective mass, as indicated. The arrows mark experimental centroids [18,19]. The black triangles mark the centroids of the pictured curves.

Clearly the centroids are influenced by the choice of the effective mass. Quantitatively the effects of the effective mass and $K_0$ choices are comparable. In addition, the results suggest the possibility to obtain a functional with the $K_0$ and m* values selected such that both nuclei are equally well described.

Further analyses are underway. In upcoming studies the sum rules of the distributions should also be examined.

**CONCLUSIONS AND PERSPECTIVES**

The KIDS functional represents a flexible, convenient and robust method for direct connections between finite nuclei and infinite matter, both ways. A minimal form has been determined and explored. In the process, it was found that the isoscalar effective mass parameters of analytical EDFs like Skyrme can hardly been constrained from bulk and static nuclear properties, but they do affect significantly the dynamic response, as exemplified here by the details of the electric-dipole transition strength distribution in $^{68}$Ni and the energy of the giant monopole resonance in $^{120}$Sn. Based on preliminary results it seems feasible to find values of $K_0$ and m*/m such as to resolve the issue of the GMR energy in Sn isotopes. Further explorations are in progress.

So far, a minimal KIDS-Skyrme interaction has been considered, namely three orders of density-dependent terms but only one order of momentum dependence and no tensor force. The momentum dependence can be extended to the next order for an even richer functional without changes in the original EoS.

**Acknowledgments**
We thank Prof. Chang Ho Hyun for an attentive reading of the manusctript. Work supported by the Rare Isotope Science Project of the Institute for Basic Science funded by Ministry of Science, ICT and Future Planning and the National Research Foundation (NRF) of Korea (2013M7A1A1075764).